\documentstyle[prl,aps,psfig]{revtex}
\begin{document}
\twocolumn[\hsize\textwidth\columnwidth\hsize
           \csname @twocolumnfalse\endcsname
\title{Space-time versus particle-hole symmetry of collisions\\
in a non-local kinetic equation for dense Fermi systems}
\author{V\'aclav \v Spi\v cka and Pavel Lipavsk\'y}
\address{Institute of Physics, Academy of Sciences, Cukrovarnick\'a 10,
16200 Praha 6, Czech Republic}
\author{Klaus Morawetz}
\address{Fachbereich Physik, University Rostock, D-18055 Rostock,
Germany}
\maketitle
\begin{abstract}
Binary collisions in Fermi systems obey two fundamental symmetries
corresponding to the space and time inversion and to the interchange of
particles and holes. We show that beyond the local and instant
approximation of scattering-in and -out integrals of a kinetic equation,
only one of symmetries can be explicit while the other has to be covered
by a constraint. This constraint, derived from the optical theorem,
allows one to convert at need an explicit particle-hole symmetric form
to the space-time symmetric form and vice versa. We implement this
constraint to heavy ion reactions, where simulation algorithms require
the space-time symmetry while former theories offer kinetic equations
with the explicit particle-hole symmetry.
\end{abstract}
\vskip2pc]
Realistic simulations of heavy ion reactions in the non-relativistic
energy range have achieved a level on which basic problems with a
reliable treatment of the Boltzmann equation (BE) \cite{BG88} or of the
quantum molecular dynamics (QMD) \cite{A91} are already settled. The
field is now open for a refinement of underlying equations of motion.
Three kinds of improvements are at hand: momentum-dependent mean field
\cite{GBD87}, medium effects on binary collisions \cite{ARSKK96}, and
non-local binary collisions \cite{H81,M83,KDB96}. If implemented
separately, each kind is covered by a corresponding theory. The mean
field follows from the Landau theory of Fermi liquids \cite{D84,BM90}.
The medium effects have been derived within the many-body statistics
\cite{ARS94}. The non-local collisions are supported by the theory of
gases \cite{CC90}. In realistic studies all the three kinds should
appear together, but none of the mentioned theories covers this general
case.

The kinetic equation with the Landau-type mean-field and in-medium
non-local collisions has been derived from non-equilibrium Green's
functions \cite{SLM98}. Although this kinetic equation includes all
the three kinds of corrections, it is not suited for the numerical
simulations because of the symmetry reason. In the BE or QMD
simulations each single collision event plays a dual role of the
scattering-in and -out so that the -in process naturally emerges as 
the space-time mirror of the -out process. Conversely, only theories
offering non-local collisions with the space-time symmetry can be
implemented. In the kinetic equation of \cite{SLM98} the non-local
corrections to scattering-out are not the space-time mirror of the
scattering-in as a consequence of the particle-hole symmetry of 
non-equilibrium Green's functions. Indeed, the particle-hole 
symmetry demands that hole-hole collisions (scattering-out) have 
the same scattering rate, including non-local corrections, as the 
particle-particle collisions (scattering-in). Accordingly, an 
{\em explicit} particle-hole symmetry excludes an {\em explicit} 
space-time symmetry. 
 
The explicit space-time symmetry is absent not only in \cite{SLM98} 
but in any theoretical treatment of non-local scattering integrals
\cite{B69,L89,NTL91,BKKS96}. The problem has not been recognized,
however, due to non-transparent forms of non-local corrections (in fact
not aimed for simulations). We anticipate that in all mentioned approaches 
the collisions obey both symmetries, the particle-hole symmetry is
explicit while the space-time symmetry is hidden in the optical theorem. 
In this letter we derive a constraint with which one can interchange 
roles of symmetries making the space-time one explicit and the 
particle-hole one hidden. Although we focus on simulations of heavy 
ion reactions, the discussed problem exceeds merits of the nuclear 
physics as the symmetry of non-local collisions in Fermi systems is 
a general question. The key step of our approach can be easily 
translated to any formalism favoured in other fields of physics 
because it is a straightforward substitution based on the optical 
theorem.

To demonstrate the conflict between the particle-hole and the space-time
symmetry, we discuss the system of hard spheres used in model studies of
heavy ion reactions without \cite{H81} and with \cite{M83,KDB96} the
Pauli blocking. Without the Pauli blocking the kinetic equation proposed
by Enskog has non-local scattering integrals in which two colliding
particles are displaced by their diameter $D$, \cite{CC90}
\begin{equation}
{\partial f_1\over\partial t}\!+\!{k\over m}
{\partial f_1\over\partial r}=
\int\! dP\left[f_3f_4^--f_1f_2^+\right].
\label{e1}
\end{equation}
Here $f_1=f(k,r,t)$, $f_2^+=f(p,r+\Delta_2,t), f_3=f(k-q,r,t), f_4^-=
f(p+q,r-\Delta_2,t)$, $dP$ is the differential cross section, and
$\Delta_2={q\over|q|}D$ is the displacement. The essential part of the
Enskog approach is a symmetry between the scattering-in and -out. If
the scattering-out describes a process $k,p\to k\!-\!q,
p\!+\!q$, its conjugate scattering-in process, $k\!-\!q,p\!+\!q\to k,p$
is its space and time mirror. Both symmetry operations are necessary.
The time inversion reverses the process but it also flips directions of
momenta which are then flipped back by space inversion. Due to the
space inversion, the displacements in the scattering-in and -out
integrals have opposite signs.

Let us try to introduce the Pauli blocking of final states by hole
distributions. From phase-space trajectories one expects that the final
states have the same displacements as the initial ones, therefore the
right hand side of (\ref{e1}) should be extended as
\begin{equation}
\!\int\! dP\left[(1\!-\!f_1)(1\!-\!f_2^-)f_3f_4^--
f_1f_2^+(1\!-\!f_3)(1\!-\!f_4^+)\right].
\label{e3}
\end{equation}
This scattering integral violates the particle-hole symmetry. Indeed,
with the interchange of particles and holes the scattering-in and -out
interchange their roles, therefore the directions of displacements 
appear reversed. This demonstrates the conflict between space-time 
and particle-hole symmetries.

The conflict follows from incompatible concepts of the scattering-out
in the classical and the quantum statistics. One has to
recognize that a realistic collision has a finite duration and to
compare these two concepts in the time picture. Within the classical
picture of the scattering-out, the instant of the transition is
attributed to the beginning of the collision when the particle enters
particle-particle correlations. Related to the instant of transition,
the collision process thus happens in future. A different picture is
found within the quantum statistical approaches where the causal
perturbative expansion excludes any dependence on processes which
are to happen in future. The scattering-out is hence represented by
a collision of two holes in which one of the holes annihilates the
scattered-out particle.  Since the particle is annihilated at the end
of the hole-hole correlation, the collision process happens in the
past, i.e., before the instant of the transition. The classical 
particle faces its partner of the scattering-out in front of it, 
while the quantum hole leaves its hole-partner behind. 

To break the ties of causality, we need the anti-causal expansion. The 
link between the causal and the anti-causal pictures of the binary 
collision is covered by the optical theorem for the two-particle 
T-matrix. We thus use the optical theorem to derive a constraint under 
which both pictures of the collision are equivalent. The implementation 
of the optical theorem can be done on a rather general level. The 
scattering-in and -out integrals result from anticommutators, $\{.,.\}$, 
of the Kadanoff and Baym equation \cite{D84,BM90}
\begin{eqnarray}
\{G^>,\Sigma^<\}\!-\!\{G^<,\Sigma^>\}&=&
\left\{G^>,G^>\circ T_R(G^<G^<)T_A\right\}
\nonumber\\
&-&\left\{G^<,G^<\circ T_R(G^>G^>)T_A\right\}.
\label{e7}
\end{eqnarray}
Here, $G^<$ and $G^>$ are particle and hole correlation functions, the
$\circ$ denotes that $G^{>,<}$ closes one loop of the two-particle
function on its right hand side. The T-matrices and pairs of
single-particle correlation functions, $(GG)$, obey standard
two-particle operator products. With the first and second terms
interpreted as the scattering-in and \mbox{-out}, respectively, the
particle-hole symmetry is apparent from the interchange
$>\longleftrightarrow <$.

Of the essential importance is that operators in (\ref{e7}) do not
commute, what reflects the non-locality of collisions. The quasiclassical
approximation of (\ref{e7}) thus  yields non-local scattering integrals
\cite{SLM98}. Naturally, non-local corrections to scattering integrals
are determined by the order of operators. Formula (\ref{e7}) has the
causal structure expressed by the order `retarded--correlation--advanced'
of the two-particle functions. Our aim is to rearrange (\ref{e7}) so that
it will include the scattering-out in terms of the anti-causal expansion.
In the optical theorem,\footnote{
The theorem represents two alternative expressions of the anti-hermitian
part of the T-matrix, $M=i(T_R-T_A)$. From the ladder equation in the
differential form, $T_{R,A}^{-1}=V-{\cal G}_{R,A}$, with
${\cal G}_{R,A}$ given by the time cut of the spectral function
${\cal A}=i({\cal G}_R-{\cal G}_A)$, follows $i(T_R^{-1}-
T_A^{-1})=-{\cal A}$. Multiplying $M$ by $T_R^{-1}$ one finds $T_R^{-1}M
=i-iT_R^{-1}T_A$. Substituting $T_R^{-1}=T_A^{-1}+i{\cal A}$ one finds
the causal side of the optical theorem $M=T_R{\cal A}T_A$. The
anti-causal side, $M=T_A{\cal A}T_R$, follows from
$MT_R^{-1}=i-iT_AT_R^{-1}$ and the same substitution.}
\begin{equation}
T_R{\cal A}T_A=T_A{\cal A}T_R,
\label{e8}
\end{equation}
on the left hand side, the time cuts of the retarded and the advanced
functions restrict internal time integrals to the past while on the
right hand side to the future. As we will see, the change in the order
of operators results in the space-time inversion of non-local corrections.

Further progress depends on the approximation of the T-matrix. Let us 
first approximate the T-matrix by Bruckner's reaction matrix for which 
the two-particle spectral function, ${\cal A}=(G^>G^>)$, allows only for
unoccupied states. The optical theorem then yields
\begin{eqnarray}
\{G^>,\Sigma^<\}\!-\!\{G^<,\Sigma^>\}&=&
\left\{G^>,G^>\circ T_R(G^<G^<)T_A\right\}
\nonumber\\
&-&\left\{G^<,G^<\circ T_A(G^>G^>)T_R\right\}.
\label{e9}
\end{eqnarray}
The first and second terms correspond to the scattering-in and -out
integrals, respectively. Expression (\ref{e9}) has the desired explicit
space-time symmetry contrary to explicit particle-hole symmetry of
(\ref{e7}).

To see it in detail, we substitute (\ref{e9}) or (\ref{e7}) into the
Kadanoff and Baym equation and apply the quasiclassical and quasiparticle
approximations \cite{D84,BM90}. Keeping internal gradients of the
scattering integrals to the linear order, one arrives at the kinetic
equation
\begin{eqnarray}
{\partial f_1\over\partial t}&+&{\partial\varepsilon_1\over\partial k}
{\partial f_1\over\partial r}-{\partial\varepsilon_1\over\partial r}
{\partial f_1\over\partial k}
\nonumber\\
&=&\int dP^-\ f_3^-f_4^-\bigl(1-f_1\bigr)\bigl(1-f_2^-\bigr)
\nonumber\\
&-&\int dP^\pm\ \bigl(1-f_3^\pm\bigr)\bigl(1-f_4^\pm\bigr)f_1f_2^\pm,
%&-&\int dP^+\ \bigl(1-f_3^+\bigr)\bigl(1-f_4^+\bigr)f_1f_2^+,
\label{e11}
\end{eqnarray}
where superscripts $+$ and $-$ correspond to (\ref{e9}) and (\ref{e7}),
respectively, and denote signs of non-local corrections:
$f_1\equiv f(k,r,t)$, $f_2^\pm\equiv f(p,r\!\pm\!\Delta_2,t)$,
$f_3^\pm\equiv f(k\!\pm\!q\!\pm\!\Delta_K,r\!\pm\!\Delta_3,t\!\pm\!
\Delta_t)$, and $f_4^\pm\equiv f(p\!+\!q\!\pm\!\Delta_K,r\!\pm\!
\Delta_4,t\!\pm\!\Delta_t)$. The $dP^+$ is obtained from $dP^-$ by the 
flip of signs of all $\Delta$'s involved. More details about the 
limiting procedure and the differential cross section $dP^-$ the reader 
can find in \cite{SLM98}. One can see that with superscript $-$ the 
scattering-out is the particle-hole mirror of the scattering-in, while 
with $+$ it is the space-time mirror.

All non-local corrections are given by derivatives of the scattering
phase shift \mbox{$\phi={\rm Im\ ln}T_R(\Omega,k,p,q,t,r)$}
\cite{SLM98,L89,NTL91},
\begin{eqnarray}
\Delta_t&=&{\partial\phi\over\partial\Omega},\ \ \ \ \ \ \ \ \
\Delta_K={1\over 2}{\partial\phi\over\partial r},\ \ \ \ \ \ \ \
\Delta_3=-{\partial\phi\over\partial k},
\nonumber\\
\Delta_2&=&{\partial\phi\over\partial p}-{\partial\phi\over\partial q}-
{\partial\phi\over\partial k},\ \ \ \ \ \ \ \ \ \
\Delta_4=-{\partial\phi\over\partial k}-{\partial\phi\over\partial q}.
\label{e6}
\end{eqnarray}
The collision is of finite duration $\Delta_t$ during which particles
can gain momentum $\Delta_K$ due to the medium effect on the collision.
Three displacements $\Delta_{2,3,4}$ correspond to initial and final
positions of two colliding particles, one position being fixed at $r$.
The kinetic equation is justified to the linear terms in the $\Delta$'s
which directly corresponds to the linear approximation in gradients.

Without many algebraic details it can be indicated why the change of the
causal picture into the anti-causal one results in the flipped signs of
all non-local corrections. Writing the T-matrices as products of the
amplitude and the phase, $T_R=|T|{\rm e}^{i\phi}$ and
$T_A=|T|{\rm e}^{-i\phi}$, one can see that the interchange of retarded
and advanced T-matrices merely flips the sign of the phase shift $\phi$.
As all $\Delta$'s depend linearly on $\phi$, the non-local corrections
to the anti-causal scattering-out have signs reversed compared to the
causal one.

Equivalency of both forms of kinetic equation (\ref{e11}) can be expressed 
by a simple constraint. Expanding the scattering-out to the linear terms,
\begin{equation}
\int\! dP^\pm\ \bigl(1\!-\!f_3^\pm\bigr)\bigl(1\!-\!f_4^\pm
\bigr)f_1f_2^\pm=I_{\rm out}\pm \sum_\alpha 
C_{\rm out}^\alpha\Delta_\alpha,
\label{con2}
\end{equation}
where $I_{\rm out}$ is the local and instant scattering-out of the BE 
and $C_{\rm out}$ 
are coefficients of non-local corrections, one finds that amplitudes of 
anti-causal and causal coefficients equal while signs are opposite. 
Since both forms are equivalent, the sum of non-local corrections to the 
scattering-out vanishes,
\begin{equation}
\sum_\alpha C_{\rm out}^\alpha\Delta_\alpha=0.
\label{con2a}
\end{equation}
Perhaps, it is not necessary to remind that constraint (\ref{con2a}) is
a direct consequence of the optical theorem (\ref{e8}). The constraint
gives us a freedom to express the non-local scattering-out as the 
particle-hole or the space-time mirror of the scattering-in, or if of 
interest, to suppress the non-local corrections to the scattering-out 
at all.

\begin{figure} 
\psfig{figure=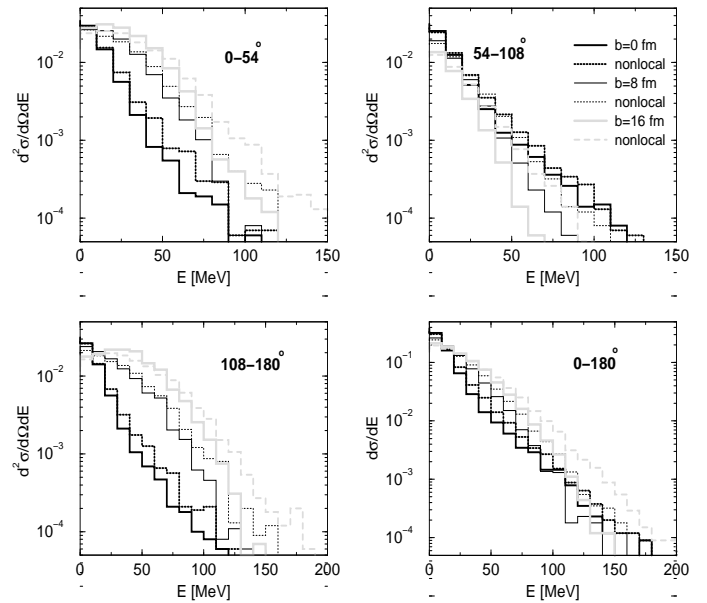,width=9cm,height=8cm,angle=-90}
\caption{The angular resolved and absolute (lower right) BUU proton
spectra for central collision of $^{209}$Bi$\rightarrow$$^{136}$Xe at 
$55$~MeV/A with and without non-local corrections. Three different impact 
parameter are chosen correspondingly. The local BUU code has been kindly 
provided us by W.~Bauer.}
\label{f1}
\end{figure}

Now the theory is ready for implementation into the BE or QMD simulations.
In Fig.~\ref{f1} we present the results of the BE simulation obtained with
a modified BUU code of \cite{KDB96}. The non-local corrections are 
evaluated from formulas (\ref{e6}) and the Paris potential with no 
adjustable parameter. In accordance with (\ref{con2a}), for the 
scattering-out we have used the standard local approximation of the cross 
section and the local Pauli blocking. In Fig.~\ref{f1} one can see that 
the production of high energetic protons is enhanced due to non-local 
corrections. The effect of non-local collisions is more pronounced for 
larger impact parameter and for forward/backward angles where it 
achieves experimentally detectable values. A comparison with
experiment and more technical details will be published elsewhere 
\cite{MLSCN98} together with the non-local QMD simulations. 

The conflict between the space-time and particle-hole symmetries deserves
a discussion from a more general point of view. The above pragmatic 
solution aimed for simulations is not the full story because constraint 
(\ref{con2a}) holds only for one particular case, the approximation of 
the T-matrix by Bruckner's reaction matrix. Moreover, when the 
particle-particle T-matrix $T$, evaluated from ${\cal A}=(G^>G^>)$, 
is used in the scattering-in, the strict demand of the particle-hole
symmetry implies to use hole-hole T-matrix $T'$, evaluated from 
${\cal A}'=(G^<G^<)$, for the scattering-out. This strict particle-hole 
symmetry cannot be demanded, however, within the Bruckner approximation
because the conservation laws require equal scattering rates of -in 
and -out processes but for $|T|\not=|T'|$ the rates result different. 

The full particle-hole symmetry is obtained within the Galitskii-Feynman 
approximation \cite{D84,BM90} for which the two-particle spectral 
function ${\cal A}=(G^>G^>)-(G^<G^<)$ corresponds to the Pauli blocking 
of internal states $(1-f)(1-f)-ff=1-f-f$. From the optical theorem 
(\ref{e8}) one then finds a modification of (\ref{con2})
\begin{equation}
\int\! dP^\pm\ \bigl(1-f_3^\pm-f_4^\pm\bigr)f_1f_2^\pm=
I_{1\!-\!f\!-\!f}\pm \sum_\alpha C_{1\!-\!f\!-\!f}^\alpha\Delta_\alpha,
\label{con2m}
\end{equation}
with a constraint $\sum_\alpha C_{1\!-\!f\!-\!f}^\alpha\Delta_\alpha=0$. 
This allows
us to write two equivalent kinetic equations with either space-time or
particle-hole symmetric scattering integrals
\begin{eqnarray}
{\partial f_1\over\partial t}&+&{\partial\varepsilon_1\over\partial k}
{\partial f_1\over\partial r}-{\partial\varepsilon_1\over\partial r}
{\partial f_1\over\partial k}
\nonumber\\
&=&\int dP^-\ f_3^-f_4^-\bigl(1-f_1-f_2^-\bigr)
\nonumber\\
&-&\int dP^\pm\ \bigl(1-f_3^\pm-f_4^\pm\bigr)f_1f_2^\pm .
\label{e11m}
\end{eqnarray}
Again, only one of the symmetries can be explicit, nevertheless, the
constraint guarantees that both symmetries are fulfilled at the same
time.

The kinetic equation (\ref{e11m}) is not suitable for numerical 
treatment by recent codes, because of negative scattering 
rates for $1-f-f<0$. Recent codes deal exclusively with the Pauli 
blocking of form $(1-f)(1-f)$ which keep the scattering rates positive. 
Numerical studies of non-local corrections are thus limited to 
the Bruckner approximation. 

We note that on the level of analytic theories, two equivalent kinetic 
equations (\ref{e11m}) settle the conflict between the space-time and 
particle-hole symmetries. The space-time symmetric form is more convenient
for studies of the conservation laws because it allows one to employ
symmetry operations developed within the classical theory of gases
\cite{CC90}. Since the non-local corrections to the scattering integrals
make the conservation laws appreciably more complex, see \cite{SLM98},
the explicit space-time symmetry offers a vital simplification.

In summary, for non-local collisions the scattering integrals cannot
have explicit space-time and particle-hole symmetries at the same time.
If one of the symmetries is made explicit, the other is covered by
a constraint following from the optical theorem. Under the condition
that the Pauli blocking of internal and final states are consistent,
this constraint simply states that the non-local corrections to the
scattering-out vanish. From the practical point of view, this constraint
gives us a freedom to select the explicit symmetry of non-local
corrections in the scattering-out at convenience, or to suppress them.
With recent simulation algorithms, tractable non-local corrections are
restricted to Bruckner's approximation of the T-matrix. We do not find 
this limitation discouraging since Bruckner's theory has been quite 
successful in studies of the equilibrium nuclear matter.

This work was supported from the GAASCR, Nr.~A1010806, the GACR, 
Nos.~202960098 and 202960021, the BMBF, Nr.~06R0884, and the
Max-Planck-Society.

\end{document}